\documentclass[12pt]{article}
\usepackage{a4wide,epsfig}

\def\nn{\nonumber}
\def\vep{\varepsilon}
\def\vk{{\bf k}_{\perp}}

\def\vd{{\bf \Delta}_\perp}
\newcommand\be{\begin{equation}}
\def\bea{\begin{eqnarray}}
\def\eea{\end{eqnarray}}

\def\mev{\,{\rm MeV}}
\def\gev{\,{\rm GeV}}

\def\eea{\end{eqnarray}}

\newcommand{\psla}{p\kern-1.0ex/}
\newcommand{\qsla}{q\kern-1.1ex/}
\newcommand{\esla}{\epsilon\kern-1.0ex/}
\newcommand{\Ksla}{K\kern-1.0ex/}

\def\als{\alpha_s}
\def\ale{\alpha_{\rm em}}

\newcommand{\lsim}{\raisebox{-4pt}{$\,\stackrel{\textstyle
                                                         <}{\sim}\,$}}

\newlength{\abstwidth}
\begin{document}
\sloppy
\renewcommand{\arraystretch}{1.5}
\pagestyle{empty}
\begin{flushright}
WU-B 00--09 \\
May, 30 2000\\
hep-ph/0005318
\end{flushright}
\vspace{\fill}
\begin{center}
{\LARGE\bf Large momentum transfer electroproduction 
       of mesons}\\[10mm]
{\Large Han Wen Huang and Peter Kroll} \\[3mm]
{\small\it Fachbereich Physik, Universit\"at Wuppertal,
42097 Wuppertal, Germany}\\[1mm]
\end{center}
\vspace{\fill}

\begin{center}
{\bf Abstract}\\[2ex]
\begin{minipage}{\abstwidth}
Assuming the proton's light-cone wave function to be dominated by
small parton virtualities and small intrinsic transverse momenta, we
show that the electroproduction amplitudes at large momentum transfer
factorize into parton-level subprocess amplitudes and form factors
representing $1/x$-moments of skewed parton distributions. On the
basis of a wave function overlap model for the form factors we
present detailed predictions for the electroproduction cross
sections. We also comment on large momentum transfer photoproduction.
\\
\end{minipage}
\end{center}
\vspace{\fill}
\clearpage
\pagestyle{plain}
\setcounter{page}{1} 
\section{Introduction}
The interest in hard exclusive reactions has recently been renewed in
the context of skewed parton distributions (SPDs) \cite{mue98,ji97,rad97}. 
The SPDs, defined as hadronic matrix elements of bilocal products of
quark or gluon field operators, are hybrid objects which combine
properties of form factors and ordinary parton distributions. In fact
reduction formulas reveal the close connection of these quantities. It
has been shown that, at large photon virtuality, $Q^2$, and small
momentum transfer, deeply virtual Compton scattering (DVCS)
\cite{rad97,ji98} and deeply virtual electroproduction of mesons (DVEM) 
\cite{rad96,CFS} factorize into hard photon-parton scattering and
SPDs describing the soft coupling between partons and hadrons. 
DVEM is dominated by longitudinally polarized photons for
$Q^2\to\infty$; the cross section for transversally polarized 
photons is suppressed by $1/Q^2$. Complementary to the large $Q^2$
region is the large momentum transfer region (small $Q^2$). In 
this kinematical region Compton scattering off protons factorizes into
a hard parton-level subprocess and a soft proton matrix element that
is described by new form factors \cite{DFJK1}. These form 
factors represent $1/x$-moments of SPDs at large momentum
transfer. Based on light-cone wave function overlaps as a model for
the SPDs, detailed predictions for cross sections and polarization
observables for real and virtual Compton scattering have been achieved
in Refs.\ \cite{DFJK1,rad98a,DFJK2}. 

\begin{figure}[t]
\begin{center}
\psfig{file=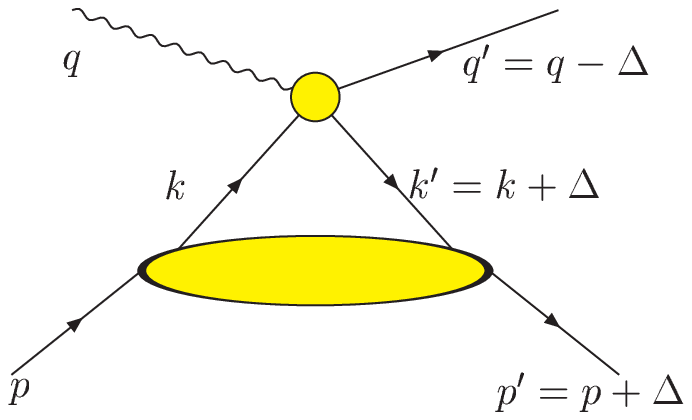, bb=155 550 356 670, height=4cm}
\end{center}
\caption{The handbag-type diagram for electroproduction of
mesons. The large blob represents a sum over all spectator configuration.
$k$ and $k'$ denote the momenta of the active partons. The small blob
stands for meson electroproduction off partons.}
\label{fig:1}
\end{figure}

Here, in this work, we are going to apply the soft mechanism proposed
in Refs.\ \cite{DFJK1,rad98a}
to electroproduction of flavor neutral pseudoscalar
($P=$ $\pi^0$, $\eta$, $\eta'$) and longitudinally polarized vector
($V=$ $\rho^0$, $\omega$, $\phi$) mesons. We will show that all
arguments given in Ref.\ \cite{DFJK1} in order to establish
factorization of Compton scattering, apply here too. I.e.\ provided
the virtualities of the partons and their intrinsic transverse
momenta, defined with respect to their parent proton's momentum, are
restricted by the proton's wave function, the dominant contribution to 
electroproduction is generated from the handbag-type diagram shown in 
Fig.\ \ref{fig:1}. It factorizes into meson electroproduction off
partons and soft proton matrix elements described by the same type of
form factors as appear in Compton scattering. It is shown in Ref.\
\cite{DFJK1} that, at large momentum transfer, there is one parton 
with a large virtuality that couples to the meson and forces the
exchange of at least one hard gluon. We, therefore, follow the concept    
used in the calculation of DVEM \cite{rad96,CFS} and treat meson
electroproduction off partons to leading-twist, lowest order
perturbative QCD. A purely soft mechanism for large momentum transfer
electroproduction of mesons, i.e.\ a soft overlap of the three
light-cone wave functions for the hadrons involved is not possible
\cite{DFJK1}. It is to be stressed 
however that the soft mechanism is not dominant for asymptotically
large momentum transfer. In this limit the hard perturbative
mechanism, for which all partons participate in the hard process,
provides the leading contribution \cite{bro80} and the soft one merely
represents a power correction. In this respect factorization of the
soft mechanism is not on the same footing as the one, say, for DVEM,
where the factorising diagrams  are dominant for asymptotically large
photon virtuality, and where factorization can be proven to hold in
all orders of perturbation theory. The soft mechanism applies 
to photoproduction of mesons as well. However, the contributions from the
hadronic component of the photon seem to dominate these processes for
values of energy and momentum transfer accessible in current experiments. 

It is also important to realize that the soft mechanism is
complementary to the perturbative one, and both the contributions have  
to be taken into account in principle. However, recent developments,
initiated by the CLEO measurements \cite{CLEO} of the $\pi\gamma$
transition form factor and its theoretical analysis, e.g.\
\cite{JKR,mus97,KR}, revealed that soft contributions play an
important role in hard exclusive reactions at experimentally
accessible momentum transfer which is of the order of a few
GeV. Indeed, in the case of the electromagnetic form factor of the
proton, the perturbative contribution has been shown to be small as
compared to experiment \cite{ber95,bol96} provided the end-point
regions, where one of the parton momentum fractions tends to zero, and
where perturbative QCD is not applicable \cite{isg}, are sufficiently
suppressed. This can be achieved by employing the modified
perturbative approach \cite{bot89} in which the transverse degrees of
freedom and Sudakov suppressions are taken into account. 

The soft contribution to large momentum transfer Compton scattering
evaluated along the same lines as for the electromagnetic form
factors, is in reasonable agreement with experiment \cite{DFJK1,rad98a}.  
The perturbative contribution, on the other hand, has only been
calculated to leading-twist accuracy \cite{kron91} and is way below the
Compton data \cite{shu79} unless strongly asymmetric, i.e.\ end-point
concentrated distribution amplitudes are used. These give, however,
results for which the bulk of the contribution is accumulated in the
soft end-point regions where the assumptions of leading-twist
perturbative calculations break down. Even if asymmetric
distribution amplitudes are utilized one obtains a perturbative
contribution to Compton scattering that likely amounts to less than
$10\%$ of the data for momentum transfer in the region of a few GeV 
\cite{kron91}; the onset of the perturbative regime is expected to be 
above 10 GeV. The calculation of the leading-twist perturbative
contribution to photoproduction of mesons has been attempted by Farrar
et al. \cite{far}. The results are at drastic variance with experiment
\cite{shu79,and76} and need verification since the method for the
numerical integrations used by Farrar et al.\ is questionable and
known to fail in Compton scattering. On account of experience with
electromagnetic form factors and Compton scattering, we will assume
that the soft contribution to electroproduction of mesons are much
larger than the perturbative ones for momentum transfers of the order
of a few GeV and that the onset of the perturbative regime is beyond
10 GeV. There is still another contribution to electroproduction: it
has two active partons, the photon couples to one of them while, by
insertion of a hard gluon, the other one generates the vector
meson. This contribution has the topology of the so-called cat's ears
diagrams. It has been discussed in Ref.\ \cite{DFJK1} that, in the
large momentum transfer region, large virtualities or intrinsic
transverse momenta occur in these diagrams forcing the exchange of
additional hard gluons. It is reasonable to assume that the magnitude
of the cat's ears contribution is between the soft and the
perturbative ones.  

The paper is organized as follows: In Sect.\ 2 we will present the
derivation of the soft mechanism. Next we will discuss the necessary
phenomenological input that parameterizes the soft hadronic matrix
elements (Sect.\ 3). In Sect.\ 4 we will comment on the case of
photoproduction and then present our results for electroproduction
of mesons (Sect.\ 5). In Sect. 6 we present our summary.
\section{The soft mechanism}
\label{sec:sur}
\begin{figure}[t]
\begin{center}
\psfig{file=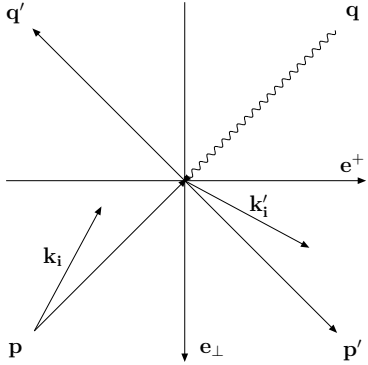, bb=205 470 405 670, height=5cm}
\end{center}
\caption{Light-cone plus and transverse components of hadron, photon
and parton momenta in the symmetric frame.} 
\label{fig:2}
\end{figure}

We are interested in electroproduction of mesons in the
kinematical region where the Mandelstam variables $s =(p+q)^2$, $-t
=-\Delta^2$ and $-u =-(p-q')^2$ are large on a hadronic scale, 
$\Lambda$, of order 1 GeV. $Q^2$ is not considered as a large
scale. Therefore the limit $Q^2\to 0$, the case of photoproduction, is
included in the following. The calculation of soft contributions to
the process of interest can be performed in full analogy to the
case of Compton scattering; all its steps can be adopted
straightforwardly. We can, therefore, restrict ourselves to an outline
of the calculation; for details we refer to \cite{DFJK1}. The
process amplitude is evaluated from the handbag-type diagram shown in
Fig.\ \ref{fig:1} where also the four-momenta are defined (as usual 
$Q^2=-q^2$). We work in a symmetric frame where the transverse momenta
of the incoming and outgoing protons are treated in a symmetric way
(see Fig.\ \ref{fig:2}) 
\begin{equation}
p=\left[p^+,\frac{m^2-t/4}{2p^+},-\frac{1}{2}\vd\right]\,,\hspace{12mm}
p^{\prime}=\left[p^+,\frac{m^2-t/4}{2p^+},\frac{1}{2}\vd\right]\,,
\end{equation}
where $m$ is the mass of the proton. The plus and minus light-cone
components of the momentum transfer are zero in this frame
$(\Delta^+=\Delta^-=0)$ and therefore $t=-\vd^2$. 
The chief advantage of the symmetric frame is that the skewedness 
parameter, defined by
\begin{equation}
\zeta= - \frac{\Delta^+}{p^+} = 1 - \frac{p^{'+}}{p^+} \,,
\end{equation}
is zero. In order to specify the frame fully we further impose
$p_3+q_3=0$. It coincides with c.m. frame for photoproduction with the
3-axis along ${\bf p}+ {\bf p}^{\prime}$.

The parton momenta are denoted by $k_i$ and $k^{\prime}_i$. 
They are characterized by the usual momentum fractions
\begin{equation}
x_i=k^+_i/p^+, \hspace{12mm} x^{\prime}_i=k^{\prime +}_i/p^{\prime +},
\end{equation}
and the transverse components $\vk{}_i$ and $\vk^{\prime}{}_i$. 
Because of $\zeta=0$ in the frame we are working, $x_i=x_i^{\prime}$. 
The arguments of the light-cone wave functions are given by the
momentum fractions and the intrinsic transverse parton momenta, i.e.\
the transverse components of the parton momenta in a frame where the
transverse momentum of the parent proton is zero. By performing
appropriate (transverse) boosts one finds for the light-cone wave
function arguments of the incoming hadron  
\begin{equation}
 \tilde{x}_i=x_i\,, \hspace{12mm} 
                   \tilde{{\bf k}}_{\perp i} = \vk{}_i + x_i\,\vd/2 \,.
\end{equation}
The arguments of the light-cone wave function of the scattered proton are
$\hat{x}^{\prime}=x_i^{\prime}$ and $\hat{{\bf k}}^{\prime}_{\perp i} =
\vk^{\prime}{}_i -x^{\prime}_i\,\vd/2$. For the sake of notation, we
henceforth drop the subscripts for the active partons, i.e.\ for
those participating in the subprocess that mediates the photon-meson 
transition (see Fig.\ \ref{fig:1}).

\begin{figure}[ht]
\begin{center}
\psfig{file=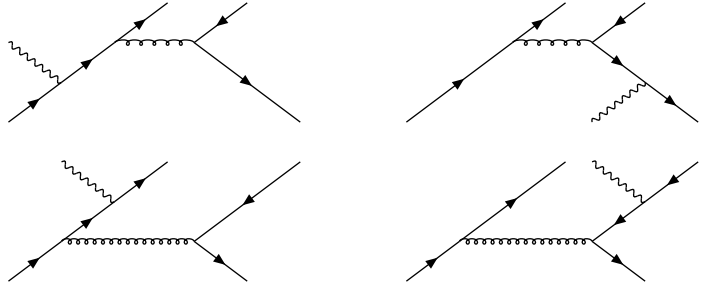, bb=110 540 510 710, height=4cm}
\end{center}
\caption{Lowest order Feynman graphs contributing to the subprocess 
$\gamma^*q\rightarrow Mq$ where $q$ is either a quark or an antiquark.
The upper quark and antiquark lines enter the meson's wave function.
The internal curly lines represent hard gluons.}
\label{fig:3}
\end{figure}

The crucial hypothesis in the soft physics approach is now that the
soft proton wave functions, i.e. the full wave functions with their 
perturbative tails removed from them, are dominated by parton
virtualities in the range $|k_i^2|, |k_i^{\prime 2}|\lsim\Lambda^2$ 
and by intrinsic transverse parton momenta satisfying 
$\tilde{{\bf k}}{}_{\perp i}^2/x_i,
\hat{{\bf k}}^{\prime 2}{}_{\perp i}/x^{\prime}_i\lsim\Lambda^2$.
\footnote{
A restriction to intrinsic transverse momenta 
$\tilde{{\bf k}}{}_{\perp i}^2\lsim\Lambda^2$ instead of 
$\tilde{{\bf k}}{}_{\perp i}^2/x_i\lsim\Lambda^2$ fails as is shown in
\cite{DFJK1}. At least one of the parton virtualities would be of 
order $\Lambda\sqrt{-t}$ and not $\Lambda^2$.} 
With the help of this hypothesis one can show \cite{DFJK1} that
the subprocess Mandelstam variables, $\hat{s}=(k+q)^2$ and
$\hat{u}=(k^{\prime}-q)^2$, are respectively equal to $s$ and $u$ up
to corrections of order $\Lambda^2(t\pm Q^2)/t$ provided $s$ and 
$-u$ are large on a hadronic scale. This implies that the poles at 
$\hat{s}\simeq 0$ and $\hat{u}\simeq 0$ appearing in the lowest order
Feynman graphs that contribute to the subprocess $\gamma^*
q\rightarrow Mq$ (see Fig.\ \ref{fig:3}) are avoided and, hence, 
the pole contributions can be neglected.  
The physical situation is that of a hard parton-level subprocess 
($\hat{s},-t,-\hat{u}\gg\Lambda^2$) and the soft emission and 
reabsorption of a parton by the proton described by a soft proton 
matrix element. Hence, we can write the helicity amplitude for
the process $\gamma^* p\rightarrow M p$ as
\begin{eqnarray}
{\cal M}^{M(q)}_{\mu^{\prime}\nu^{\prime},\,\mu\nu} &=& \sum_a ee_a B_a^M\,
        \int d^4k\, \theta(k^+) \, \int\frac{d^4z}{(2\pi)^4}\,
                                               e^{ik\cdot z} \nn\\
                &\times& \left[\, \left\langle p^{\prime}\nu^{\prime}
    \left|\,T\overline{\psi}_{a\alpha}(0)\psi_{a\beta}(z)\,\right|p\nu
                                \right \rangle
    H^{M(q)\alpha\beta}_{\mu^{\prime}\mu}(k^{\prime},k)\right. \nn\\ 
      &+&\, \left.\left \langle p^{\prime}\nu^{\prime}\left|\, 
                T\overline{\psi}_{a\alpha}(z) \psi_{a\beta}(0)\,
         \right|p\nu\right\rangle H^{M(q)\alpha\beta}_{\mu^{\prime}\mu}
                  (-k,-k^{\prime}) \,\right] \,,
\label{ia}
\end{eqnarray}
where $H^{M(q)}_{\mu^{\prime}\mu}$ is the tree-level expression for the
hard scattering kernel. $\mu$ and $\mu^{\prime}$ respectively denote the 
helicities of the photon and the meson, $\nu$ and $\nu'$ those of the
protons. For the sake of legibility we label explicit helicities only
by their signs, e.g.\ we write $+$, $-$ instead of $+1/2$, $-1/2$ for 
fermions. The helicities are defined in the $\gamma^* p$ c.m.\ frame
which is convenient for phenomenological applications and facilitates
comparison with other results. On the other hand, the symmetric frame 
is adapted to discuss the reaction mechanism. The sum runs over quark 
flavors $a$, $e_a$ being the electric charge of quark $a$ in units of 
the positron charge $e$ and $B^M_a$ denotes the meson's flavor wave 
function. The first term in (\ref{ia}) corresponds to the case where 
the incoming parton in the subprocess is a quark, the second term 
corresponds to an incoming antiquark. For the production of flavor 
neutral vector mesons gluons have to be considered as active partons 
too. We will discuss this contribution separately below.

Since the subprocess is dominated by a large scale, we can approximate the 
momenta $k,k^{\prime}$ of the active partons in the subprocess
as being on-shell, collinear with their parent hadrons 
\begin{equation}
k \simeq \left[k^+ ,-\frac{t}{8k^+ },-\frac{1}{2}\vd\right]\,, \hspace{10mm} 
k^{\prime} \simeq \left[k^+ ,-\frac{t}{8k^+ },\frac{1}{2}\vd\right]\,.
\label{kj}
\end{equation}
The integration over $k^-$ and $\vk$ in (\ref{ia}) can then
be performed explicitly leaving an integral $\int dk^+ \int dz^-$ and
forcing the relative distance of fields in the matrix elements on the
light cone, $z\to \bar{z}=[0,z^-,\bf{0}_{\perp}]$. After this the time
ordering of the fields can be dropped \cite{die98a}. 

\begin{figure}[bt]
\begin{center}
\psfig{file=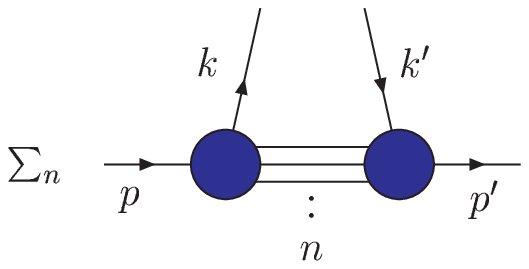, bb=225 580 380 660, height=3cm}
\end{center}
\caption{Parton picture of the soft proton matrix element.}
\label{fig:4}
\end{figure}

The proton matrix element can be viewed as the amplitude for a proton 
with momentum $p$ emitting the active parton with momentum $k$ and a 
number of on-shell spectators times the corresponding conjugated 
amplitude for $p^{\prime},k^{\prime} $ summed over all spectator 
configurations, see Fig.\ \ref{fig:4}. This corresponds to inserting a 
complete set of intermediate states between quark and antiquark fields 
in (\ref{ia}). Realizing that at the proton-parton vertices one has 
large plus components but, on account of the central hypothesis of
small parton virtualities and small intrinsic transverse momenta, 
$\tilde{{\bf k}}^2_{\perp i}/x_i, \hat{{\bf
k}}^{\prime 2}_{\perp i}/x^\prime_i\lsim\Lambda^2$, one cannot form
large kinematical invariants. With this feature of the soft mechanism 
at hand one can replace the products of fields in (\ref{ia})
by
\bea
 \overline\psi{}_{\alpha}(0)\, \psi_{\beta}(\bar{z}) &\to&
 \left( \frac{1}{2k^+} \right)^2 \sum_{\lambda,\lambda'}
 \left( \overline\psi(0) \gamma^+ u(k',\lambda') \right) \nn\\
 &\times& \left( \bar{u}(k,\lambda) \gamma^+ \psi(\bar{z}) \right) \;
 \bar{u}_{\alpha}(k',\lambda') \, u_{\beta}(k,\lambda)\,,
\label{replace}
\eea
where $\lambda$ and $\lambda^{\prime}$ denote the helicities of the
active partons and $u$ their on-shell spinors. An analogous
replacement is possible for the product 
$\overline{\psi}_{a\alpha}(\bar{z})\psi_{a\beta}(0)$. In this case
antiquark spinors, $v$, appear. Due to this replacement the hard
scattering kernels in (\ref{ia}) are multiplied with the spinors for 
on-shell (anti)quarks
\begin{equation}
{\cal H}^{M(q)}_{\mu^{\prime}\lambda^{\prime},\,\mu\lambda} \,=\, 
        \bar{u}(k^{\prime}, \lambda^{\prime})\,
          H^{M(q)}_{\mu',\,\mu} (k^{\prime},k)\, u(k,\lambda)\,,
\end{equation}
which guarantees electromagnetic gauge invariance of our result. The 
charge conjugation properties of Dirac matrices and spinors relate the 
subprocess amplitudes involving antiquarks to the quark amplitudes
\begin{equation}
\bar{v}(k,\lambda)\, H^{M(q)}_{\mu'\mu}(-k,-k^{\prime})\, 
                   v(k^{\prime} ,\lambda^{\prime})\,=\, 
        \kappa_M\, {\cal H}^{M(q)}_{\mu^{\prime}\lambda^{\prime},\,\mu\lambda}\,,
\end{equation}
where $\kappa_V=-1$ for vector mesons and $\kappa_P=+1$ for
pseudoscalar ones. The replacement (\ref{replace}) reveals that the
plus components of the non-local currents dominate the proton matrix
element and that the operators in the matrix elements are in fact the
same as those of the leading-twist parton distributions occurring
in deep-inelastic lepton-nucleon scattering, DVCS or DVEM. This is a 
nontrivial dynamical feature of large momentum transfer Compton
scattering and electroproduction of mesons, given that, in contrast to  
the deeply virtual reactions, not only the plus components of the
parton momenta but also their minus and transverse components are
large now. 

As mentioned above we follow the concept used in the calculation of
DVEM \cite{rad96,CFS,man98} and treat the formation of the helicity zero
mesons ($\mu'=0$) to leading-twist, lowest order perturbative QCD 
(cf.\ Fig.\ \ref{fig:3}). In combination with the disregard of quark
masses this formation mechanism leads to conservation of quark
helicity in the subprocess, $\lambda^{\prime}=\lambda$. This feature 
and properties of massless spinors allow to simplify the expression 
(\ref{ia}) further, and to arrive at
\begin{eqnarray} 
   {\cal M}^{M(q)}_{0\nu^{\prime},\,\mu\nu} &=& \frac{1}{4} \sum_{\lambda}  \,
   \sum_a e e_a B^M_a\, \int \frac{dk^+}{k^+}\,\theta{(k^+)} \,
   \int {d z^-\over 2\pi}\, e^{i\, k^+ z^-}\, 
             {\cal H}^{M(q)}_{0\lambda, \,\mu\lambda} \nonumber \\
           &\times& \left[ \langle p'\nu '|\,
     \overline\psi{}_{a}(0)\, \gamma^+\,\psi_{a}(\bar{z}) +\kappa_M
     \overline\psi{}_{a}(\bar{z})\, \gamma^+\,\psi_{a}(0)
     \,|p\nu\rangle \right. \nonumber \\
         &+& \left. \lambda\, \langle p'\nu '|\,
     \overline\psi{}_{a}(0)\, \gamma^+\gamma_5\,\psi_{a}(\bar{z}) -\kappa_M 
     \overline\psi{}_{a}(\bar{z})\, 
\gamma^+\gamma_5\,\psi_{a}(0)\, |p\nu\rangle \right]  \,.
\label{final}
\end{eqnarray}
Following \cite{DFJK1}, we take $k^+=p^+$, i.e.\ the light-cone
fractions $x=x^{\prime}=1$ in the hard scattering which is in line
with the requirement to have no hard parton directly coupling to the
protons. Admittedly, the global factor $1/k^+$ in (\ref{final}) cannot
be plainly associated with either the hard scattering or the soft matrix
element. We therefore choose to keep $k^+=xp^+$ for this factor. We
can now pull out the hard scattering amplitude from the  
integrals and use a form factor decomposition for the integrated
proton matrix element \cite{DFJK1,rad98a} 
\begin{eqnarray}
&&\int^1_0\frac{dx}{x}p^+\int\frac{dz^-}{2\pi}e^{ixp^+z^-}
\left\langle p^{\prime}\nu^{\prime}\left|
       \overline{\psi}_a(0)\gamma^+\psi_a(\bar{z}) + \kappa_M
       \overline{\psi}_a(\bar{z})\gamma^+\psi_a(0)
                                 \right|p\nu \right\rangle \nn\\
&&= R^{M a}_V(t)\,\bar{u}(p^{\prime},\nu^{\prime})\gamma^+u(p,\nu)
+R^{M a}_T(t)\frac{i}{2m}\, \bar{u}(p^{\prime},\nu^{\prime})\sigma^{+\beta}
\Delta_{\beta}u(p,\nu)\,, \nn\\
&&\int^1_0\frac{dx}{x}p^+\int\frac{dz^-}{2\pi}e^{ixp^+z^-}
              \left\langle p^{\prime}\nu^{\prime}\left|
           \overline{\psi}_a(0)\gamma^+\gamma_5 \psi_a(\bar{z}) - \kappa_M
          \overline{\psi}_a(\bar{z})\gamma^+\gamma_5\psi_a(0)
                 \right|p\nu \right\rangle \nn\\
&&= R^{M a}_A(t)\,
\bar{u}(p^{\prime},\nu^{\prime})\gamma^+\gamma_5 u(p,\nu)\,.
\label{spd}
\end{eqnarray}
$R_V^{M a}$, $R_T^{M a}$ and $R_A^{M a}$ are new form factors, 
depending on the type of the meson, $V$ or $P$, and on the flavor of 
the active quark. As the definition (\ref{spd}) reveals they are 
$1/x$-moments of SPDs at zero skewedness. The link-operator needed to 
render the definition of the SPDs gauge invariant, is not displayed in 
(\ref{spd}), i.e.\ we assume the use of a light-cone gauge combined 
with an appropriate choice for the integration path which reduces the 
link operator to unity. Due to time reversal invariance the form
factors are real functions. The form factor $R_T^{Ma}$ is controlled 
by higher-twist dynamics and is expected to be suppressed by $m^2/t$ 
as compared with $R_V^{M a}$ \cite{DFJK2}. Since the calculation of 
the soft contributions is only accurate up to corrections in 
$\Lambda^2/t$, $R_T^{Ma}$ is to be omitted for consistency. Hence, we 
can only calculate the amplitudes conserving the proton helicity. 
Explicitly they read 
\bea
{\cal M}^{M(q)}_{0+,\,\mu +}(s,t)&=&\frac{e}{2}\, 
                 \left\{{\cal H}^{M(q)}_{0+,\,\mu +}(s,t)\,
                         [R_V^{M}(t)+R_A^{M}(t)] \right.\nn\\
             && \hspace{3mm} + \left. {\cal H}^{M(q)}_{0-,\,\mu -}(s,t)\,
                         [R_V^{M}(t)-R_A^{M}(t)]\right\}\,,
\label{amp}
\eea
where the form factors specific to the process
$\gamma^*p\rightarrow Mp$ are defined as
\begin{equation}
R_{V,A}^M(t)=\sum_{a}\, e_a\, B^M_a R_{V,A}^{M a}(t)\,.
\label{ffdecomp}
\end{equation}
{}From parity invariance one has 
${\cal M}^{M(q)}_{0\nu,\,\mu\nu}=-\kappa_M(-1)^{\mu}
{\cal M}^{M(q)}_{0-\nu,-\mu -\nu}$ and an analogous equation for the  
parton-level amplitudes ${\cal H}^{M(q)}_{0\lambda^{\prime},\,\mu\lambda}$.
The amplitudes for longitudinally polarized photons simplify as a 
consequence of parity invariance: the vector form factor, $R^M_V$, 
contributes only in the case of vector meson production while the
axial vector form factor, $R^M_A$, contributes in the case of
pseudoscalar mesons. This is analogous to DVEM. For transversally
polarized photons, on the other hand, both form factors contribute. 

Let us now turn to the calculation of the parton-level amplitudes. 
The mesons are described by their valence Fock components and, for
a longitudinally polarized vector meson, we write the 
corresponding matrix element in the usual way as 
\begin{eqnarray}
\left\langle V, q^{\prime} \left |\overline{\psi}(x)\gamma_{\mu}\psi(y)
         \right | 0 \right\rangle = q^{\prime}_{\mu}f_V\int_0^1d\tau\phi_V
       (\tau)\, e^{iq^{\prime}\cdot (\tau x+\bar{\tau}y)}\,,
\label{vda}
\end{eqnarray}
where the proportionality between the meson's polarization vector and
its momentum, $q^\prime$, for longitudinally polarized vector mesons is
employed. For pseudoscalar mesons we have:
\begin{eqnarray}
\left\langle P, q^{\prime} \left | \overline{\psi}(x)\gamma_5\gamma_{\mu}
\psi(y)\right |0\right\rangle = i q^{\prime}_{\mu}f_P\int_0^1d\tau\phi_P
       (\tau)\, e^{iq^{\prime}\cdot (\tau x+\bar{\tau}y)}\,.
\label{pda}
\end{eqnarray}
The meson masses are ignored. $\tau$ is the fraction of the meson's
momentum the valence quark in the meson carries. The momentum fraction
of the antiquark is $\bar{\tau} =1-\tau$. $f_{M}$ is the meson's decay
constant and $\phi_M$ its distribution amplitude which is normalized as
\begin{equation}
\int_0^1 d\tau \phi_M(\tau)=1\,.
\end{equation}
The definitions (\ref{vda}) and (\ref{pda}) are 
equivalent to the other frequently used ones \cite{bro80} 
$(\rlap/{q}^{\prime}/\sqrt{2}) f_V \phi_V/(2\sqrt{2N_c})$ and 
$(\rlap/{q}^{\prime}\gamma_5/\sqrt{2}) f_P \phi_P/(2\sqrt{2N_c})$.
The color factor $1/\sqrt{N_c}$ (where $N_c$ denotes the number of
colors) is not displayed in (\ref{vda}) and (\ref{pda}); it is taken
into account in the parton-level amplitudes explicitly.

Working out the Feynman graphs shown in Fig.\ \ref{fig:3}, one finds
for the parton-level amplitudes
\begin{equation}
{\cal H}^{M(q)}_{0+,\,\mu +}\,=\, 2\pi\alpha_s(\mu_R)f_M \frac{C_F}{N_c}\,
                          \int_0^1d\tau\phi_M(\tau)f^{(q)}_{\mu}(\tau)\,,
\label{subamp}
\end{equation}
where 
\begin{eqnarray} 
f^{(q)}_{+}(\tau)&=& \frac{\sqrt{-2t}}{s+Q^2}\,
            \left\{\frac{(s+Q^2)(\tau s+Q^2)-\bar{\tau}uQ^2}
                        {\bar{\tau} s(\tau t-\bar{\tau}Q^2)} 
                    +\frac{(s+Q^2)(\tau s-Q^2)-\bar{\tau}uQ^2}
                        {{\tau}u (\bar{\tau}t-{\tau}Q^2)}
                                            \right\}\,, \nn\\
f^{(q)}_{-}(\tau)&=&- \bar{\tau} \frac{\sqrt{-2t}}{s+Q^2}\,
             \left\{\frac{u}{\bar{\tau} (\tau t-\bar{\tau}Q^2)}
                  +\frac{s}{\tau (\bar{\tau}t-{\tau}Q^2)}
                                            \right\}\,, \nn\\
f^{(q)}_0(\tau)&=& \frac{2Q\sqrt{-su}}{s+Q^2}
            \left\{\frac{u}{s (\tau t-\bar{\tau}Q^2)}
                  +\frac{s+Q^2 + \bar{\tau} u}
       {\tau u(\bar{\tau}t-{\tau}Q^2)}\right\}\,.
\label{eq:fi}
\end{eqnarray}
$C_F=(N_c^2-1)/(2N_c)$ is the usual SU(3) color factor. Parity
invariance fixes the amplitudes with negative quark helicities. For
the scale of the parton-level amplitudes we choose $\mu_R=s/4$ which
is roughly the average of the gluon and quark virtualities in the hard
process. 

In principle the amplitudes (\ref{subamp}) hold for all values of $t$
and $Q^2$ provided the internal quark and gluon virtualities are
sufficiently large. In the limit of either $Q^2\rightarrow 0$ or 
$t\rightarrow 0$ the amplitudes (\ref{subamp}) simplify strongly. In 
the case of photoproduction we find
\begin{eqnarray}
{\cal H}^{M(q)}_{0+,++}&=& -2\pi\alpha_s(\mu_R)f_M \frac{C_F}{N_c}\,
             \langle 1/\tau\rangle_M\, \frac{\sqrt{-2t}}{u}\,,\nn\\
{\cal H}^{M(q)}_{0+,-+}&=&\phantom{-}
                2\pi\alpha_s(\mu_R)f_M \frac{C_F}{N_c}\,
              \langle 1/\tau \rangle_M\, \frac{\sqrt{-2t}}{s}\,,
\label{subphoto}
\end{eqnarray}
and ${\cal H}^{M(q)}_{0+,0+}=0$ ($\propto Q$ for $Q^2\rightarrow
0$). In deriving (\ref{subphoto}) we made use of the symmetry of the
distribution amplitude for the mesons of interest under the
interchange $\tau \leftrightarrow \bar{\tau}$,
$\phi_{M}(\tau)=\phi_{M}(\bar{\tau})$. One observes that only the moment
\begin{equation}
\langle1/\tau\rangle_M=\int^1_0d\tau\frac{\phi_M(\tau)}{\tau}
\label{moment}
\end{equation}
contributes. In the limit of large $Q^2$ and small $-t$, the case of
DVEM, the amplitude for longitudinally polarized photons, ${\cal
H}^{M(q)}_{0+,0+}$, also becomes proportional to the $1/\tau$ moment,
cf.\ for instance \cite{man98}, while terms $\propto 1/\tau^2$, 
$1/\bar{\tau}^2$ in the other two amplitudes signal the break-down of 
factorization for transversally photons \cite{man00}. Inserting the
parton-level amplitudes (\ref{subamp}) into (\ref{amp}), one obtains
the final expressions for the helicity amplitudes. 

For flavor-neutral vector meson there is a complication which we now
have to discuss, namely gluons have to be considered as active partons
as well. Again this situation is similar to DVEM
\cite{rad96,CFS,man98}. In the kinematical region of $\Lambda^2 \ll Q^2
\ll s$, characteristic of the HERA experiments, the gluon contribution
even dominates \cite{bro94,mar96}. We start the calculation of the
gluon contribution from an expression similar to (\ref{ia})
\bea
{\cal M}^{V(g)}&=&\sum_a e e_a B_a^V 
           \int d^4k\theta(k^+)\int\frac{d^4z}{(2\pi)^4}\, e^{ik\cdot z}\,\nn\\ 
    &\times&\langle p^{\prime}|T A^{\rho b}(0)A^{\rho' b'}(z) |p \rangle\,
     H^{V(g)}_{\rho\rho' b b'}(k^{\prime},k)\,,
\label{glue-amp}
\eea
where $k$ and $k^\prime$ denote the momenta of the on-shell gluons in
the symmetric frame, see (\ref{kj}). $A_{\rho b}$ is the gluon field
with color $b$. For the sake of legibility we do not display helicity
labels in this equation. The proton matrix element is only non-zero
if $b=b^\prime$. With this in mind we omit color labels in the
following for convenience.

Now, we have to repeat all steps of the derivation of the quark
contribution. As there, the use of the approximation (\ref{kj}) forces
the relative distance of the fields in the proton matrix elements to
the light cone, $z\to \bar{z}$. The important point is now the use of
light-cone gauge, $n\cdot A=0$ (where $n=\left[0,1,{\bf 0}_\perp
\right]$) which allows to express the gluon field by an integral over
the field strength tensor $G_{\nu\mu}$ \cite{rad97,rad96,kog70}
\begin{equation}
    A_\nu(\bar{z};n) = n^\mu \int_0^\infty d\sigma e^{-\vep\sigma}\,
    G_{\nu\mu}(\bar{z}+\sigma n)\,.
\end{equation}
(The limit $\vep\to 0$ is understood.)
With the help of arguments similar to those leading to (\ref{replace})
we can replace the products of fields appearing in (\ref{glue-amp}) by
\begin{eqnarray}
 A^\rho(0) A^{\rho^\prime}(\bar{z}) &=& \sum_{\lambda,\lambda^\prime=\pm 1}
                  \epsilon^\rho (k,\lambda) \epsilon^{*\rho^\prime}
                                               (k^\prime,\lambda^\prime)
                   \int d\sigma\,
                   d\sigma^\prime\, e^{-\vep\sigma-\vep^\prime\sigma^\prime} 
                                                              \nonumber\\
                 &\times& G_{\mu +}(\sigma^\prime n)        
                       G_{\mu^\prime +}(\bar{z}+\sigma n)
                   \epsilon^{*\mu}(k,\lambda)
                  \epsilon^{\mu^\prime}(k^\prime,\lambda^\prime)\,.
\label{glue-replace}
\end{eqnarray}
In the symmetric frame the polarization vectors of the on-shell gluons
read
\begin{equation}
\epsilon (k,\lambda)=[0,\epsilon^-,\epsilon_\perp(\lambda)]\,,
\hspace{1cm} 
\epsilon (k^\prime,\lambda^\prime)=[0,\epsilon^\prime{}^-,
                                     \epsilon_\perp(\lambda^\prime)]\,,
\end{equation}
where ${\bf \epsilon}_\perp(\pm 1) = \mp (1,\pm i)/\sqrt{2}$. The minus
components need not be specified since they do not contribute in
light-cone gauge.

The hard scattering kernels appearing in (\ref{glue-amp}) are
contracted by the first set of polarization vectors in (\ref{glue-replace})
which leads to gauge invariant parton-level amplitudes
\begin{equation}
    {\cal H}_{0 \lambda^\prime,\, \mu\lambda}^{V(g)} = 
            \epsilon^{*\rho^\prime}(k^\prime,\lambda^\prime)
                          H_{\rho^\prime \rho}^{V(g)}(k^\prime,k)
                               \epsilon^\rho (k,\lambda)
\end{equation}
Gluon helicity flip ($\lambda =- \lambda^\prime$) is suppressed in the
proton matrix element at large $-t$ since two units of orbital
angular momentum are required in order to avoid helicity flips of the
proton ($\nu=-\nu^\prime$). Thus, matrix elements involving helicity flips
of the gluons are suppressed at least as $\propto m^2/t$ and will be
omitted. This argument is of importance only for longitudinally
polarized photons because, for $\mu=\pm 1$, the parton-level amplitudes 
${\cal H}_{0\lambda,\, \mu -\lambda}$ come out to zero  in any case. 
For $\lambda=\lambda^\prime$ one may decompose the last part of
(\ref{glue-replace}) into an unpolarized and a polarized gluon
contribution
\bea
\lefteqn{G_{\mu +}(\sigma^\prime n) G_{\mu^\prime +}(\bar{z}+\sigma n)
             \epsilon^{*\mu}(k,\lambda) 
                       \epsilon^{\mu^\prime}(k^\prime,\lambda)=} \nn\\
  && \hspace{10mm} \frac12\, G_{\mu +}(\sigma^\prime n) 
                        G_{\mu^\prime +}(\bar{z}+\sigma n)
         \left[\, g^{\mu\mu^\prime}_\perp + 
                          \lambda\, {\cal P}^{\mu\mu^\prime}\, \right]\,,
\eea
where $g^{11}_\perp = g^{22}_\perp = 1$ and ${\cal P}^{12} = - {\cal
P}^{21} = i$ while all other components of these tensors are zero. As
for the case of quarks, see (\ref{spd}), we introduce a form factor
decomposition for the proton matrix elements of the field strength
tensors
\bea
\lefteqn{\int_0^1 dx\, p^+ \int \frac{dz^-}{2\pi}\, e^{ixp^+z^-} \int_0^\infty
    d\sigma\,d\sigma^\prime e^{-\vep\sigma-\vep^\prime\sigma^\prime}} \nn\\
   && \hspace{10mm} \times\, \langle p^\prime,\nu^\prime|\, G_{\mu +}(\sigma^\prime n)\,
                  G_{\mu^\prime +}(\bar{z}+\sigma n)\, \,|p,\nu\rangle 
                g^{\mu\mu^\prime}_\perp  \nn\\
&& \hspace{10mm} = \frac{\bar{u}(p^\prime,\nu^\prime) \gamma_+ 
                                               u(p,\nu)}{2p_+}\, R^g_V(t)
           + \frac{i}{2m} \frac{\bar{u}(p^\prime,\nu^\prime) 
                        \sigma_{+\nu}\Delta^\nu u(p,\nu)} 
              {2p_+}\, R_T^g(t)\,. \hspace{1cm}  
\eea
and 
\bea
\lefteqn{\int_0^1 dx\, p^+ \int \frac{dz^-}{2\pi}\, e^{ixp^+z^-} \int_0^\infty
d\sigma\,d\sigma^\prime e^{-\vep\sigma-\vep^\prime\sigma^\prime}} \nn\\   
  && \hspace{10mm} \times\, \langle p^\prime,\nu^\prime|\, G_{\mu +}(\sigma^\prime n)\,
               G_{\mu^\prime +} (\bar{z}+\sigma n)\,|p,\nu\rangle 
               \, {\cal P}^{\mu\mu^\prime} \nn\\
&& \hspace{10mm}= \frac{\bar{u}(p^\prime,\nu^\prime) \gamma_+\gamma_5
                            u(p,\nu)}{2p_+}\,  R^g_A(t)\,. \hspace{4cm} 
\eea
The form factors are related to SPDs at zero skewedness (${\cal
F}^g_{\zeta=0}$, ${\cal K}^g_{\zeta=0}$ and ${\cal G}^g_{\zeta=0}$),
e.g.\
\be
 R^g_V(t) = \int_0^1 \frac{dx}{x^2} {\cal F}^g_{\zeta=0} (x,t)\,.
\end{equation}
Since the forward limits of ${\cal F}^g_{\zeta}$ and ${\cal
G}^g_{\zeta}$ are defined in such a way that 
\be 
 xg(x) = {\cal F}^g_{\zeta=0}(x,t=0)\,, \quad\quad 
               x\Delta g(x) = {\cal G}^g_{\zeta=0}(x,t=0)\,,
\end{equation}
one may still call these form factors $1/x$-moments of
SPDs. Neglecting, as in the case of quarks, $R_T^g$ we finally arrive at
the helicity amplitudes for the gluon contribution 
\bea 
{\cal M}_{0+,\,\mu +}^{V(g)}(s,t) &=& \frac{e}{2} \left[ {\cal H}^{V(g)}_{01,\,\mu
             1}(s,t) \left( R^{V(g)}_V(t) +  R^{V(g)}_A(t)\right)
                                                 \right.    \nn\\
            &+& \left. {\cal H}^{V(g)}_{0-1,\,\mu -1}(s,t)  
             \left( R^{V(g)}_V(t) -  R^{V(g)}_A(t)\right) \right]\,,
\label{gluonic-amplitudes}
\eea
where 
\be
R^{Vg}_{V,A} = \sum_{a} e_a B^V_a R^g_{V,A}\,.
\label{gluonff}
\end{equation}
Because of parity invariance 
$ {\cal H}^{V(g)}_{0-1,-\mu -1} = (-1)^\mu\; {\cal H}^{V(g)}_{01,\mu 1}$ 
the form factor $R^{V(g)}_A$ does not contribute to the amplitudes for
longitudinal photons.

\begin{figure}[bt]
\begin{center}
\psfig{file=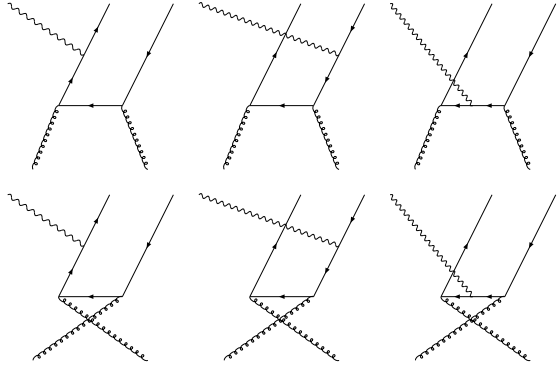, bb=90 360 520 655, height=5cm}
\end{center}
\caption{Lowest order Feynman graphs for the subprocess $\gamma^* g\to
Mg$.}
\label{fig:5}
\end{figure}
The parton-level amplitudes, to be evaluated from the six lowest order
Feynman graphs shown in Fig.\ \ref{fig:5}, read
\begin{equation}
{\cal H}^{V(g)}_{0+,\,\mu +}=\frac{2\pi\alpha_s(\mu_R)}{N_c}\, 
f_V\int_0^1 d\tau\phi_V(\tau)f^{(g)}_{\mu}(\tau)\,,
\end{equation}
where 
\bea
f^{(g)}_{+}(\tau)&=& 
                    \sqrt{\frac{t}{2su}}\,\frac{Q^2}{s+Q^2}\,
                                   \frac{1}{\tau\bar{\tau}}\, 
                    \frac{t Q^2 -(s+Q^2)^2 -4\tau\bar{\tau} s u}
                         {(\bar{\tau}t-\tau Q^2)(\tau t-\bar{\tau} Q^2)}\,,\nn\\
f^{(g)}_{-}(\tau) &=& \sqrt{\frac{ut}{2s}}\,\frac{s}{s+Q^2}\,
                                \frac{1-4\tau\bar{\tau}}{\tau\bar{\tau}}\,
                               \frac{Q^2}
                     {(\bar{\tau}t-\tau Q^2)(\tau t-\bar{\tau} Q^2)}\,, \nn\\
f^{(g)}_0(\tau)&=& \frac{Q}{s+Q^2}\, \frac{1}{\tau\bar{\tau}}\,
                 \frac{2\tau\bar{\tau}[Q^2(s+Q^2) - t (s-Q^2)] - t Q^2} 
                      {(\bar{\tau}t-\tau Q^2)(\tau t-\bar{\tau} Q^2)}\,.
\label{fg}
\end{eqnarray} 
Since the gluon amplitudes only contribute to photoproduction of
flavor neutral vector mesons whose associated distribution amplitudes
are symmetric under the interchange $\tau \leftrightarrow \bar{\tau}$,
we display only the $\tau \leftrightarrow \bar{\tau}$ symmetric part 
of the amplitudes in (\ref{fg}). The quark amplitudes (\ref{eq:fi}), 
on the other hand, contribute to the production of flavored mesons too.
We, therefore, show the full quark amplitudes although we do not
discuss these cases here.

As one may see from (\ref{fg}), the gluon amplitudes vanish in the
limit $Q^2\to 0$. Finite $\gamma g \to V g$ amplitudes may be obtained
if meson masses and/or transverse momenta are taken into account (cf.\
for instance \cite{mar96,man98a}). 

The amplitudes (\ref{glue-amp}) have to be added to those given in
(\ref{amp}) for vector mesons:
\be
{\cal M}^V_{0\nu^\prime,\,\mu\nu}\,=\,
                         {\cal M}^{V(q)}_{0\nu^\prime,\,\mu\nu}\,+\,
                         {\cal M}^{V(g)}_{0\nu^\prime,\,\mu\nu}\,. 
\end{equation}
\section{The form factors and the meson distribution amplitudes}
Before we present numerical results for the observables of 
electroproduction of mesons, we have to model the new form factors. In
Eq.\ (\ref{ffdecomp}) the general composition of these new form
factors is presented in terms of the individual flavor
contributions. The explicit flavor structure of the form factors for
various mesons is given in Tab.\ \ref{tab:1}. In contrast to Compton 
scattering \cite{DFJK1} where the sum runs over all flavors, here the 
sum is over the valence quarks of the produced mesons. In other words, 
the meson selects its valence quarks from the proton. The physical 
situation is thus similar to DVEM in this respect. For vector mesons, 
also the flavor factors associated with the gluonic form factors (see
(\ref{gluonic-amplitudes}), (\ref{gluonff})) are listed in the table.
 
\begin{table}[ht]
\begin{center}
\begin{tabular}{|c|c|c|}\hline
&$R_i^M$     &$R^{Vg}_i$\\\hline
$\rho^0,\pi^0$ & $\frac{1}{\sqrt{2}}[e_uR_i^{M u}-e_dR_i^{M d}]$
                        &  $\frac{1}{\sqrt{2}}[e_u-e_d]R^g_i$\\\hline
$\omega,\eta_q$ & $\frac{1}{\sqrt{2}}[e_uR_i^{M u}+e_dR_i^{M d}]$
                        &  $\frac{1}{\sqrt{2}}[e_u+e_d]R_i^g$\\\hline
$\phi,\eta_s$ & $e_sR^{M s}_i$ & $e_sR_i^g$\\\hline
\end{tabular}
\end{center}
\caption{Flavor composition of the form factors $R_i^M(i=V,A)$ for 
pseudoscalar and vector mesons and the flavor factors for the gluonic
form factors.}
\label{tab:1}
\end{table}

$\omega-\phi$ mixing is ignored since the corresponding mixing angle
is very small. $\eta-\eta^{\prime}$ mixing, on the other hand, 
is taken into account. Following \cite{FKS1}, we work in the quark
flavor basis and write
\begin{eqnarray}
\eta&=&\cos{\phi_P} \;\eta_q - \sin{\phi_P} \;\eta_s \,,\nn\\
\eta^{\prime}&=& \sin{\phi_P} \;\eta_q + \cos{\phi_P} \;\eta_s\,.
\label{mixing}
\end{eqnarray}
$\eta_q$ is a state built from $u$ and $d$ quarks only while $\eta_s$ is
a $s\bar{s}$ state. The parameters of that $\eta-\eta^{\prime}$ mixing
scheme, the mixing angle, $\phi_P$, and the decay constants, $f_q$ and
$f_s$, of the basis states are determined in \cite{FKS1} on exploiting
the divergencies of the axial vector currents which embody the axial
vector anomaly. For the mixing angle a value of $39.2^{\circ}$ is found
in \cite{FKS1}. 

As an inspection of Eqs.\ (\ref{final}) and (\ref{spd}) reveals
the form factors $R_{V,A}^{V a}$ are exactly the same as those
appearing in Compton scattering \cite{DFJK1,rad98a}. Thus, in
principle, from a combined analysis of data on Compton scattering and
production cross section for various mesons, one may extract
information on the form factors for individual flavors from
experiment. This allows to test the soft mechanism independent of a
specific model for the form factors. 

For a numerical estimate of the form factors we use the model proposed
in Ref.\ \cite{DFJK1}. In a frame where $\Delta^+=0$ the SPDs and, hence,
the form factors can be represented as overlaps of light-cone wave
functions summed over all Fock states in close analogy to the familiar
Drell-Yan formula \cite{DY}. A detailed discussion of that overlap 
representation is given in Refs. \cite{DFJK1,DFJK3}. Each $N$-particle 
Fock state is described by a number of terms, each with its own momentum
space wave function $\Psi_{N\beta}$, where $\beta$ labels different 
spin-flavor combinations of the $N$ partons. Assuming a single Gaussian
$\tilde{k}_{i\perp}$-dependence of the soft Fock state wave functions
\begin{equation}
\Psi_{N\beta}(x_i,\tilde{k}_{\perp i})\propto \exp[-a_N^2\sum_{i=1}^N
                    \frac{\tilde{k}^2_{\perp i}}{x_i}]\,,
\label{wf}
\end{equation}
one can explicitly carry out the momentum integration in the overlap
formula. The ansatz (\ref{wf}) satisfies various theoretical
requirements \cite{chi95,bro98} and is in line with our central
hypothesis that the soft hadronic wave functions are dominated by
transverse momenta with $\tilde{k}^2_{\perp i}/x_i\le \Lambda^2$,
necessary to achieve the factorization of the 
electroproduction amplitudes into soft and hard parts
\footnote{
          The wave function (\ref{wf}), perhaps multiplied by a
          polynomial in the $x_i$, is not continuous in the end-points
          $x_i=\tilde{k}_{\perp i}=0$ ($i=1,2$ or 3). It can, however,
          be shown \cite{DFJK3} that the overlaps evaluated from such 
          wave functions, are infrared stable, i.e.\ they are not
          dominated by contributions from regions of very small $x_i$ 
          and $\tilde{k}_{\perp i}$.}.
The results of the transverse momentum integration for the vector and 
axial vector form factors are respectively related with the Fock state
contributions to the unpolarized ($q_a(x)$, $g(x)$) and polarized
($\Delta q_a(x)$, $\Delta g(x)$) parton distributions. For simplicity 
one may assume a common transverse size parameter $a_N=\hat{a}$ for
all Fock states which seems to be a reasonable approximation since, for 
large -t, the main contribution to the overlap integral is only due to 
a limited number of Fock states \cite{DFJK1}. This simplification immediately 
allows one to sum over the Fock states without specifying the 
$x_i$-dependence of the wave functions. One then arrives at the following
model for the form factors for individual flavors ($a=u,d,s$):
\begin{eqnarray}
R^{V a}_V(t)&=&\int_0^1\frac{dx}{x}\,\exp{\left[\frac{1}{2}\hat{a}^2t
        \frac{1-x}{x}\right]}\,\{q_a(x)+\bar{q}_a(x)\}\,, \nn\\
R^{P a}_V(t)&=&\int_0^1\frac{dx}{x}\,\exp{\left[\frac{1}{2}\hat{a}^2t
       \frac{1-x}{x}\right]}\,\{q_a(x)-\bar{q}_a(x)\}\,, \nn\\
R^g_V(t)&=&\int_0^1\frac{dx}{x}\,\exp{\left[\frac{1}{2}\hat{a}^2t
           \frac{1-x}{x}\right]}g(x)\,. 
\label{overlap}
\end{eqnarray}
The corresponding axial vector form factors are obtained from
(\ref{overlap}) by replacing the unpolarized parton distributions
$q_a$, $g$ with the polarized ones, $\Delta q_a$, $\Delta g$.

\begin{figure}[hbtp]
\begin{center}
\psfig{file=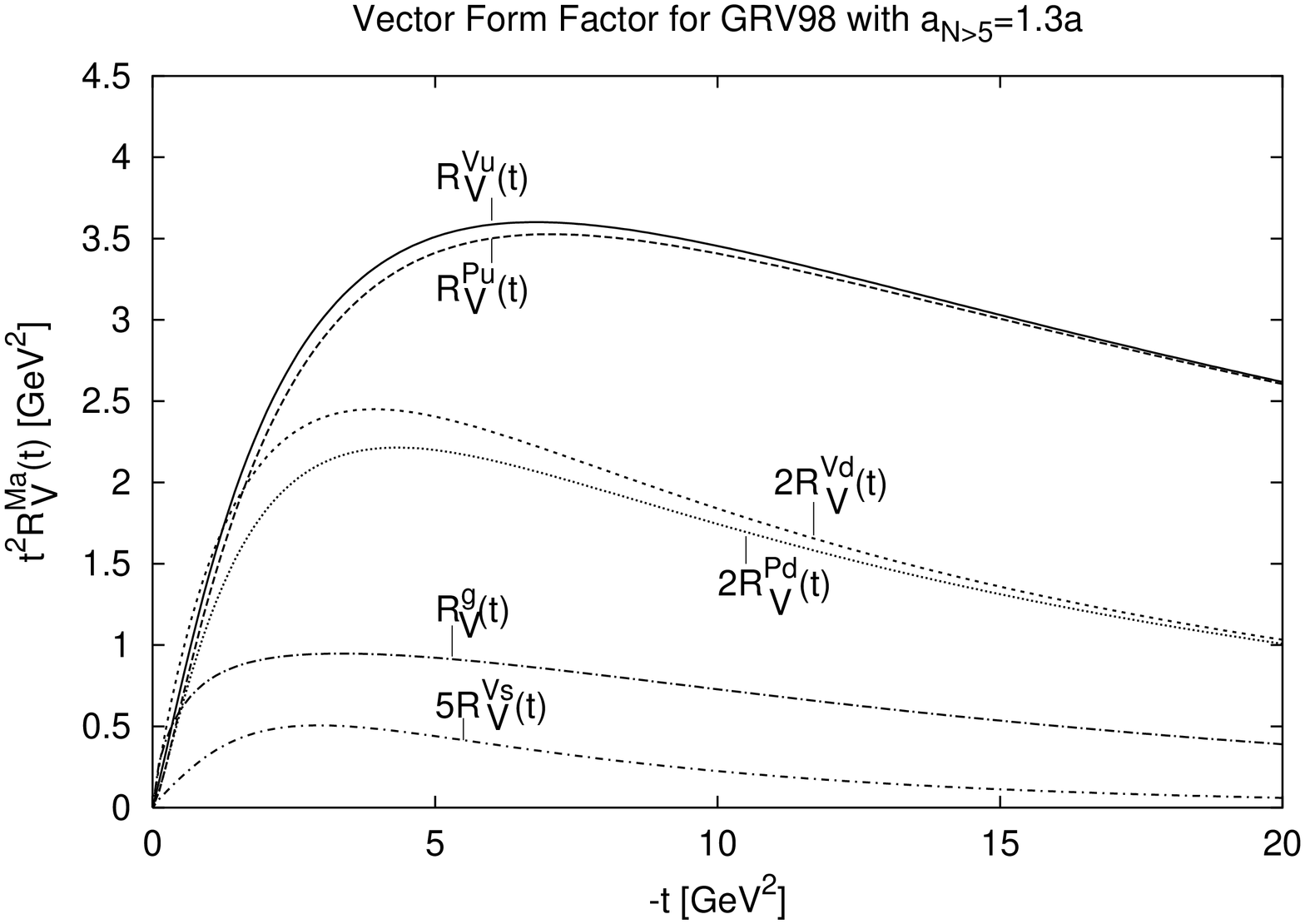,width=9.cm,angle=-90}\\
\psfig{file=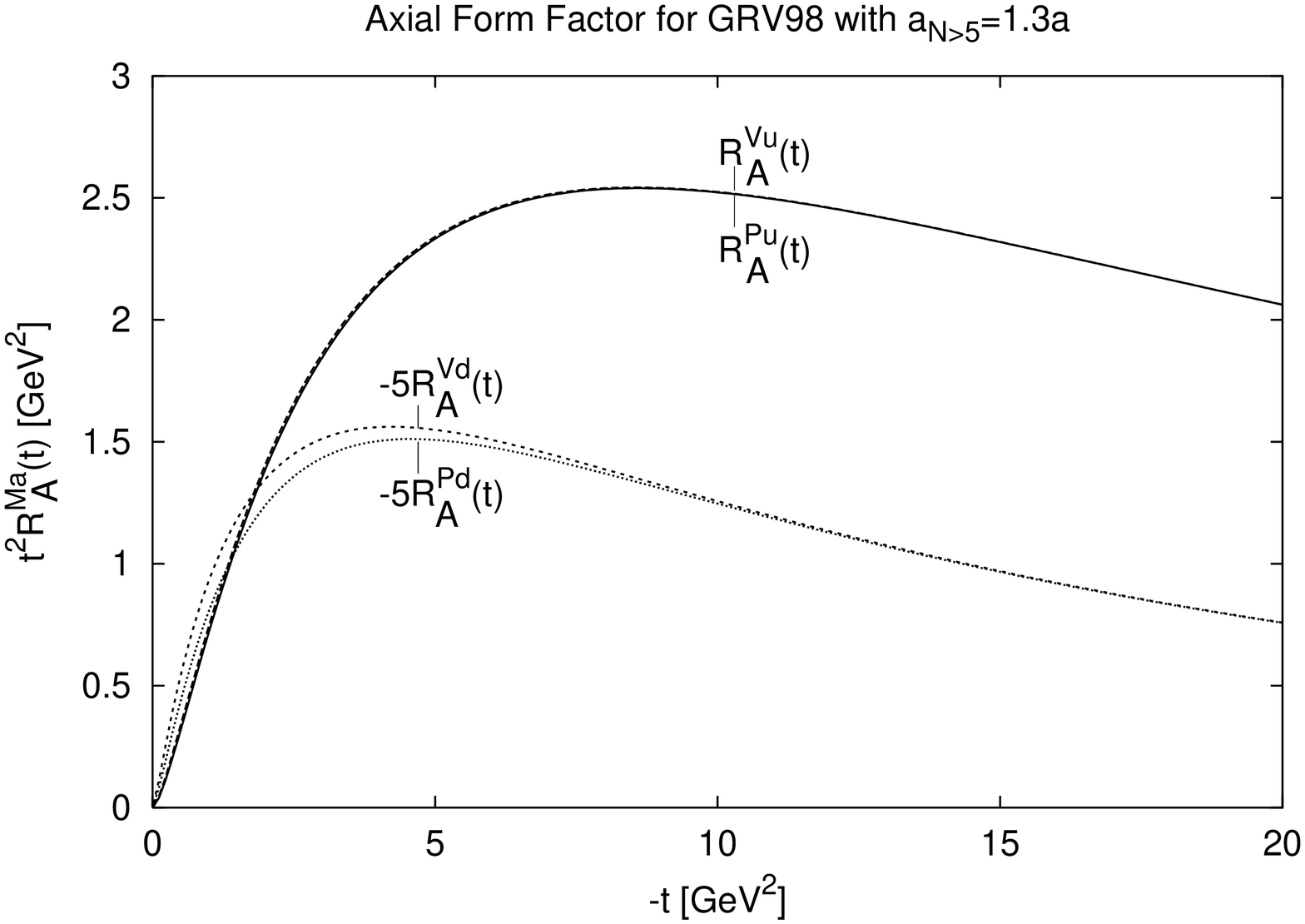,width=9.cm,angle=-90}
\end{center}
\caption{The form factors $R_V^{M a},R_A^{M a}$ and $R_V^g$, scaled by
$t^2$, as evaluated from the overlap model proposed in \cite{DFJK1}.}
\label{fig:6}
\end{figure}

As shown in \cite{DFJK1} an evaluation of these form factors from the
parton distributions of Gl\"uck et al. (GRV) \cite{GRV} (taken at a
scale of 1 GeV) and with
$\hat{a}\simeq 1\gev^{-1}$, leads to results for Compton scattering 
in fair agreement with experiment. In order to improve the model 
(\ref{overlap}) the lowest three Fock states were modeled explicitly
in \cite{DFJK1} assuming specific distribution amplitudes, e.g.\ 
\begin{equation}
\phi_{123}(x_i)=60x_1x_2x_3(1+3x_1)\,,
\label{BK-da}
\end{equation}
for the valence Fock state \cite{bol96}. The form factors
(\ref{overlap}) for the quarks are then evaluated from these three
lowest Fock states (with $a_3=a_4=a_5=0.75 \gev^{-1}$) and the 
contribution from all higher Fock states is estimated by setting 
($a_N=1.3\, a_3$ for $N>5$)
\begin{equation}
\sum_{N>5}q^{(N)}_a(x)=q_a(x)-\sum_{N=3,4,5}q_a^{(N)}(x).
\end{equation}
The $q_a(x)$ are taken from the GRV parameterization \cite{GRV} and
the three lowest Fock state contribution $q_a^{(N=3,4,5)}(x)$ are
evaluated from the light-cone wave functions. This model provides a 
good fit to Compton scattering and to the proton form factor $F_1$ (by
expressions similar to (\ref{overlap}) \cite{DFJK1}). 
In Fig.\ \ref{fig:6} numerical results for the vector and axial vector
form factors are shown. The simplifying model assumption of
unpolarized gluons and sea quarks has the consequence of a zero gluon
form factor $R^g_A$. Most of the strange form factors are therefore
zero too: 
\begin{equation}
R^{P s}_V=R^{V s}_A=R_A^{P s}=0 \,.
\label{strange}
\end{equation}
Only $R_V^{Vs}$ is non-zero, even though very small. 
The form factors for $u$ quarks are largest.
They approximately behave as $1/t^2$ in the momentum transfer region
from about 5 to $15 \gev^2$; for the other flavors and for the gluon
that range is shifted to somewhat smaller values of $-t$. With  
increasing $-t$ the form factors for $u$ and $d$ quarks gradually turn
into the soft physics asymptotic $\propto 1/t^4$, while the other form
factors decrease faster. The leading powers of $1/t$ in the
asymptotic behavior of the form factors (\ref{overlap}) follow from
the $x_i$-dependence of the model wave functions at the end points
(see for instance Eq.\ (\ref{BK-da})) \cite{DFJK1}. In the region
where the form factors drop as $1/t^4$ or faster, which is above 100
GeV$^2$, the perturbative contribution will take the lead.

The other soft physics information required in our approach is that of
the form of the meson's distribution amplitude. From analyzes of the  
pion-photon transition form factor (see for instance
\cite{JKR,mus97,KR}) it became evident that the pion's
distribution amplitude (its formal definition is given in (\ref{pda}))
is close to the asymptotic form 
\begin{equation}
\phi_{\rm AS} (\tau)=6\tau(1-\tau)\,.
\label{da}
\end{equation}
This result is supported by the instanton model \cite{pet99} and by
recent QCD sum rule studies \cite{bra99}. 
The analyzes of the $\eta$- and $\eta'$-photon transition form factors
revealed that the $\eta_q$ distribution amplitude is close to the form
(\ref{da}), too \cite{JKR,fel98}. Although the transition form factor
data are compatible with the asymptotic distribution amplitude for the 
$\eta_s$ as well, a somewhat narrower one cannot be excluded. For
vector mesons no phenomenological information is available but QCD sum rules
\cite{bal96} taught us that the distribution amplitudes for
longitudinally polarized vector mesons can also be approximated by
(\ref{da}). In order to keep matters simple we therefore choose the
form (\ref{da}) for all mesons. We expect that the uncertainties in
the predicted production cross sections due this choice do not exceed
$10 - 15 \%$. Associated with the distribution amplitude (\ref{da}) is
a value of 3 for the $1/\tau$-moment (\ref{moment}).

For the meson decay constants we use the values \cite{neu97}
\bea
f_\pi&=& 132\mev\,, \quad f_\rho= 216\mev\,, \nn\\
\quad f_\omega&=& 195\mev\,, \quad f_\phi= 237\mev\,,
\eea
and for the decay constants of the states $\eta_q$ and $\eta_s$ \cite{FKS1}
\be
f_q= 141\mev\,, \quad f_s= 177\mev\,.
\end{equation}
\section{Photoproduction of mesons}
\begin{figure}[t]
\parbox{\textwidth}{\begin{center}
\psfig{file=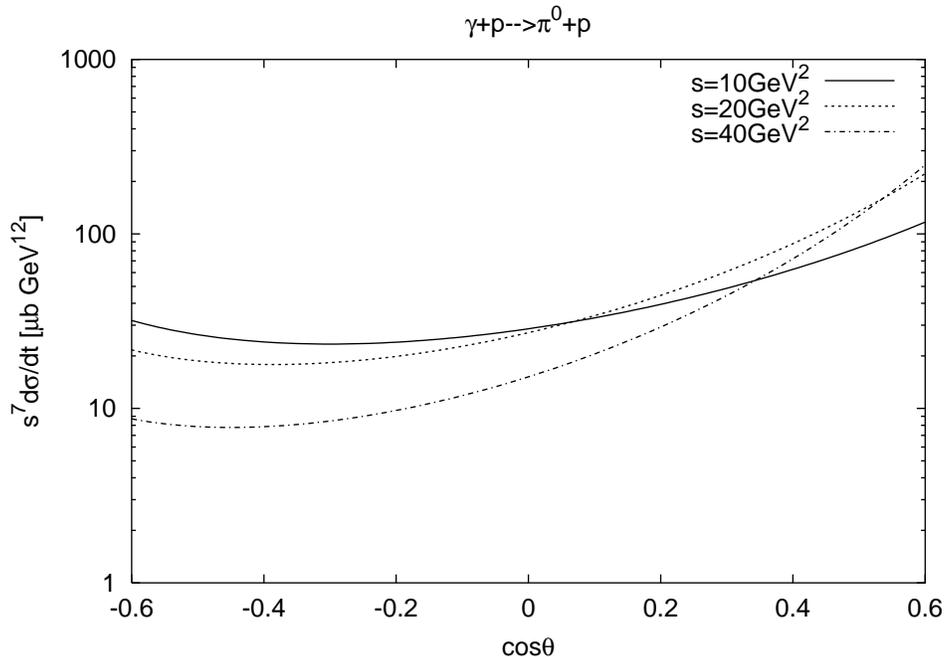,width=9cm,angle=-90}
\end{center}}
\caption{The soft physics contribution to the cross section for
photoproduction of $\pi^0$ scaled by $s^7$ versus $\cos{\theta}$, where
$\theta$ is the scattering angle in the $\gamma^* p$ c.m.\ system.} 
\label{fig:7}
\end{figure}

Using (\ref{amp}) and (\ref{subphoto}), we obtain for the
photoproduction cross section of pseudoscalar mesons
\bea
\frac{d\sigma}{dt}^P&=& \frac12 \ale \left[\pi\als(\mu_R)f_P
                 \langle 1/\tau \rangle_P \frac{C_F}{N_C}\right]^2 \nn\\
             &\times& \frac{-t}{u^2 s^4}\, 
        \left\{(s-u)^2\, (R^{P}_V(t))^2 + t^2\,(R^{P}_A(t))^2 \right\}\,.
\eea
The corresponding expressions for photoproduction of vector mesons are
a bit more complicated due to the occurrence of the gluonic contribution.
In Fig.\ \ref{fig:7} we show, as an typical example, the soft physics
contribution to the large momentum transfer photoproduction cross
section of uncharged pions as evaluated from the form factors
discussed in Sect.\ 3. We see that at fixed scattering angle the cross
section approximately exhibits the $s^7$-scaling as predicted by
dimensional counting \cite{far73}. This scaling behavior holds in the
soft physics approach as long as the form factors  
$R_{V,A}^M$ behave as $1/t^2$ (see Fig.\ \ref{fig:6}). Similar 
results are found for the production of the other mesons.

As compared to experiment (at $s\simeq 10 \gev^2$) \cite{shu79,and76} 
the soft physics contributions are too small by orders of
magnitude. This can easily be understood by evaluating the 
ratio of $\pi^0$ production and Compton cross section \cite{DFJK1}
\begin{equation}
\frac{d\sigma(\gamma p\rightarrow\pi^0p)}{d\sigma(\gamma p\rightarrow\gamma p)}
= \frac{-t}{s}\,\frac{\alpha_s^2(\mu_R)}{\alpha_{\rm em}}\,
           \frac{f_{\pi}^2<1/\tau>^2_{\pi}}{s}\,
                      c_{soft}\,,
\label{ratio}
\end{equation}
where $c_{soft}$, a ratio of form factors and kinematical factors, is
of order 1. The ratio of the two cross sections is therefore about $2
\gev^2/s$, i.e. much smaller than unity in contradiction to experiment
\cite{shu79}, where the ratio is about 50 (at $s\simeq 10 \gev^2$,
and a $\gamma^* p$ c.m.\ scattering angle, $\theta$, of $90^{\circ}$). 
Mainly responsible for the small ratio (\ref{ratio}) is the perturbative 
formation of the meson which only probes small quark-antiquark separations 
in the meson. The amplitude is, therefore, proportional to the meson's 
decay constant which, for dimensional reasons, is to be scaled 
by $\sqrt{s}$. The $\langle 1/\tau \rangle_\pi$ moment, appearing as a
consequence of the perturbative meson formation, cannot compensate
the small ratio $f_\pi/\sqrt{s}$.

The ratio (\ref{ratio}) also holds in perturbative calculations,
in the pure quark picture \cite{far} as well as in the diquark model,
a variant of the standard perturbative approach in which
diquarks are considered as quasi-elementary constituents of baryons
\cite{ber99}. The factor $c_{pert}$ may, however, be larger than
unity. Although there is no obvious enhancement in any of the 
many Feynman graphs contributing to the perturbative amplitude, the
graphs may conspire in such a way that a large value of $c_{pert}$ is
built up. In order to see whether or not this is the case,
an explicit and reliable calculation of meson production within the
perturbative approach is called for. 

The observation of a relatively large photoproduction cross section in
experiment is in line with the power law behavior in $s$ at
fixed scattering angle. $s^{-n}$-fits to the present data in the range 
$6.5 \gev^2 \lsim s \lsim 12 \gev^2$ and $50^\circ <\theta <
130^\circ$ provide \cite{shu79}:
\begin{eqnarray}
\gamma p\rightarrow\pi^0p:&&n=8.0\pm 0.1\,, \nn\\
\gamma p\rightarrow\,\gamma\; p:&&n=6.1\pm 0.3\,. \nn
\end{eqnarray}
For the $\gamma p \to (\rho^0+\omega) p$ data \cite{and76} the
statistics does not allow a meaningful determination of the power
$n$. However, $n$ seems to be larger than 7, rather compatible with 8.
For Compton scattering the power is compatible with dimensional
counting, while for $\pi^0$ production, and possibly for the 
sum of $\rho^0$ and $\omega$ production, the value of $n$ rather equals
that one observed in elastic $\pi p$ scattering \cite{bur76}
($n\simeq 8$; data are averaged over resonance-like structures)
\footnote{
     We recall that a $s^8$-scaling of the fixed-angle meson baryon cross
     section is easily accounted for by the soft physics approach in the
     relevant region of energy \cite{DFJK2}.}. 
Admittedly, the photoproduction data are rather poor and need
confirmation. Data on photoproduction of $\phi$ mesons will become
available from the TJlab soon \cite{laget} which will perhaps allow a
determination ot the power $n$ for that reaction.

Both the observations in the large  momentum transfer photoproduction
data, the large powers of $s$ at fixed scattering angle and the large
cross sections, indicates that another dynamical mechanism is at work
here. It is tempting to assign it to the hadronic component of the
photon. This proposition is supported by a vector meson
dominance (VMD) estimate of the photoproduction cross section. 
Combined with quark model ideas VMD, for instance, relates
photoproduction of $\rho^0$-mesons to elastic pion-nucleon
scattering  \cite{sak}
\begin{equation}
\frac{d\sigma}{dt}(\gamma p\rightarrow\rho^0p)=\alpha_{\rm em}
                    \,\frac{\pi f^2_\rho}{m_{\rho}^2}\, 
        \left[\frac{d\sigma}{dt}(\pi^+p \rightarrow\pi^+p)
              +\frac{d\sigma}{dt}(\pi^-p\rightarrow\pi^-p)\right]\,.
\end{equation}
This relation is satisfied by experiment within a factor of 2-3
\cite{and76}. With respect to the uncertainties arising from possible
spin effects and the poor quality of data this may be considered as
fair agreement. Thus, it seems that photoproduction of $\rho^0$ and  
$\pi^0$ -- and likely of other mesons -- is indeed controlled by the
hadronic component of the photon. In this case one would expect the 
produced vector mesons to be polarized transversally rather than 
longitudinally. Since the fixed-angle energy dependencies of the 
contributions from the hadronic component of the photon 
($\simeq s^{-8}$) and from the soft mechanism ($\simeq s^{-7}$) are 
so close, much higher energies are needed before the soft contribution 
(and/or the perturbative one) will control photoproduction of mesons. 
We, therefore, refrain 
from presenting more predictions from the soft physics approach for
photoproduction of mesons.  We stress that our approach to photoproduction
requires high energies, large momentum transfer and small values of
$|\cos{\theta}|$. If $s/-t \gg 1$ Pomeron exchange becomes dominant,
see for instance \cite{iva}.

One may wonder whether Compton scattering is also dominated by the
hadronic component of the photon. However, the analogous VMD estimate, 
with both the photons replaced by vector mesons, provides values for 
the Compton cross section that are about an order of magnitude below 
experiment \cite{shu79} at $s\simeq 10 \gev^2$. Moreover, the Compton 
cross section exhibits $s^6$-scaling and not an $s^8$ one as would be 
the case if the hadronic component of the photon dominates. Thus, 
the simplest elementary process, elastic scattering of point-like
photons from quarks, dominates Compton scattering off protons already
at rather low energies \cite{bjo69} .  
\section{Electroproduction of mesons}
Let us now discuss our results for large momentum transfer
electroproduction. As is well-known the cross section for
$ep\rightarrow epM$ can be decomposed as follows 
\begin{eqnarray}
\frac{d^4\sigma^M}{dsdQ^2dtd\varphi}&=&\frac{\ale\;s}
                                {16\pi^2E_L^2m^2Q^2(1-\epsilon)} \nn\\
                &\times& \left(\frac{d\sigma^M_T}{dt}
                               +\epsilon \frac{d\sigma^M_L}{dt}
                   +2\epsilon \cos{2\varphi} \frac{d\sigma^M_{TT}}{dt}+
                   \sqrt{2\epsilon(1+\epsilon)}\cos{\varphi} 
                           \frac{d\sigma^M_{LT}}{dt}\right)\,,
\label{epcross}
\end{eqnarray}
where $\varphi$ denotes the azimuthal angle between the hadronic and
leptonic scattering planes. $E_{L}$ is the energy of the incoming
electron in the laboratory frame and $\epsilon$ is the ratio of 
longitudinal to transverse photon flux. Details of the kinematics can 
be found for instance in Ref. \cite{kro96a}.  
The partial cross sections in (\ref{epcross}) read:\\
(i) The cross sections for transverse photons (reducing to the unpolarized cross 
section for photoproduction of mesons, i.e. for $Q^2=0$) and for longitudinal
photons,
\begin{eqnarray}
\frac{d\sigma^M_T}{dt}&=&\frac{1}{32\pi s(s+Q^2)}\,
                   \sum_{\nu',\nu}\,|{\cal M}^M_{0\,\nu',+\,\nu}|^2\,,\nn\\
\frac{d\sigma^M_L}{dt}&=&\frac{1}{32\pi s(s+Q^2)}\, 
                   \sum_{\nu',\nu}\,|{\cal M}^M_{0\,\nu',\,0\,\,\,\nu}|^2\,.
\end{eqnarray}
(ii) The transverse-transverse and longitudinal-transverse interference terms
\begin{eqnarray} 
\frac{d\sigma^M_{TT}}{dt}&=&-\frac{1}{64\pi s(s+Q^2)}\, 
       {\rm Re} \sum_{\nu',\nu} {\cal M}^{M*}_{0\,\nu',+\,\nu}\,
                            {\cal M}^{M}_{0\,\nu',-\,\nu}\,, \nn\\
\frac{d\sigma^M_{LT}}{dt}&=&-\frac{\sqrt{2}}{64\pi s(s+Q^2)}\,
             {\rm Re}\sum_{\nu',\nu} {\cal M}^{M*}_{0\,\nu',\,0\,\,\nu}\,
       \left[ {\cal M}^{M}_{0\,\nu',+\,\nu} - {\cal M}^{M}_{0\,\nu',-\,\nu}\right]\,.
\label{rtt}
\end{eqnarray}

\begin{figure}[t]
\begin{center}
\psfig{file=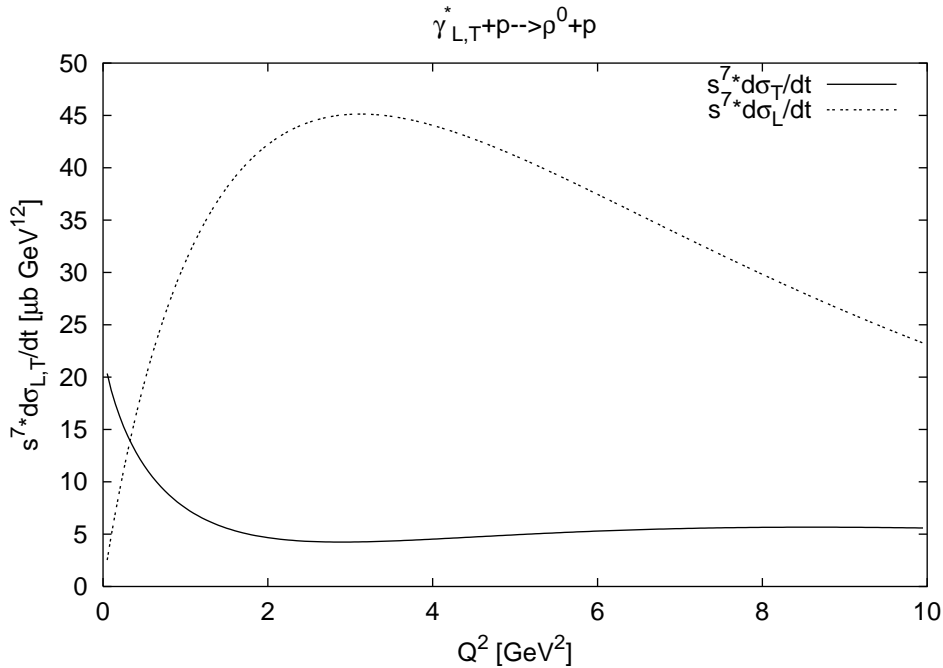,width=9cm,angle=-90}
\end{center}
\caption{The transverse and longitudinal cross sections for the
photoproduction of $\rho^0$ mesons at $s=40\,\gev^2$ and $\theta=90^{\circ}$.}
\label{fig:8}
\end{figure}
In Fig.\ \ref{fig:8} we show, as a typical example, the $\rho^0$ production
cross section for longitudinally and transversally polarized photon as
a function of $Q^2$ at a scattering angle of $90^{\circ}$.
Except for small $Q^2$ ($Q^2/s<<1$) the longitudinal cross section
\begin{eqnarray}
\frac{d\sigma^M_L}{dt}&=&
              \frac{\ale}{4N_c^2}\,\frac{[\pi\als(\mu_R)f_M C_F]^2}
              {s(s+Q^2)}\, \left\{\int^1_0d\tau\phi_M(\tau) \right. \nn\\ 
          &\times& \left[(1-\kappa_M)
    [f^{(q)}_0(\tau)R^M_V(t) + \frac{1}{C_F} f_0^{(g)}(\tau)R^{Mg}_V(t)]\right.\nn\\
&+&\left.\left. \phantom{\frac{1}{C}} \hspace{-3mm}
               (1+\kappa_M) f^{(q)}_0(\tau)R_A^M(t)\right]\right\}^2
\end{eqnarray}
dominates, i.e.\ the cross section that conserves $s$-channel
helicity. Again we have similarity to DVEM. Also 
similar is the fact that the longitudinal cross section
for the production of vector mesons is associated with the vector form
factors, $R_V^V$, while in the case of pseudoscalar mesons it is 
connected with the axial vector ones, $R_A^P$. In contrast to the
limiting cases of either $t\to 0$ or $Q^2\to 0$ the explicit form of
the mesons distribution amplitude is required in the evaluation of the
large momentum transfer electroproduction cross section. Tuning the
ratio $Q^2/t$ details of the distribution amplitude can be explored.
Even if the form factors behave  $\propto 1/t^2$ strictly,
$d\sigma/dt$ does not exhibit $s^7$-scaling. Despite of this we keep
multiplying the cross section by $s^7$ since this compensates most of
the energy dependence.
 
The hadronic component of the photon vanishes with increasing $Q^2$
rapidly, approximately as $m_V^2/(Q^2+m_V^2)$ in the amplitude. This
is for instance, clearly visible in the integrated cross section for 
$\rho^0$ production: while, at low $Q^2(\lsim 2 \gev^2)$, its energy
dependence is very similar to that of total cross sections for elastic 
hadron-hadron scattering, it is much steeper at large $Q^2$
\cite{ZEUS99}. The steep rise of the $\rho^0$ cross section is
correlated with the behaviour of the gluon SPD for $0<x^\prime \leq x
\ll 1$ which should reflect the strong increase of the gluon
distribution for $x\to 0$ \cite{rad96,mar96}.  
By virtue of the rapidly decreasing hadronic component
of the photon and the strong rise of the longitudinal cross section 
(see Fig.\ \ref{fig:8}) with increasing $Q^2$, we expect the  
soft physics approach to be applicable for photon virtualities larger
than about $2 - 3 \gev^2$ provided $s$, $-t$ and $-u$ are large. 
\begin{figure}[tp]
\begin{center}
\psfig{file=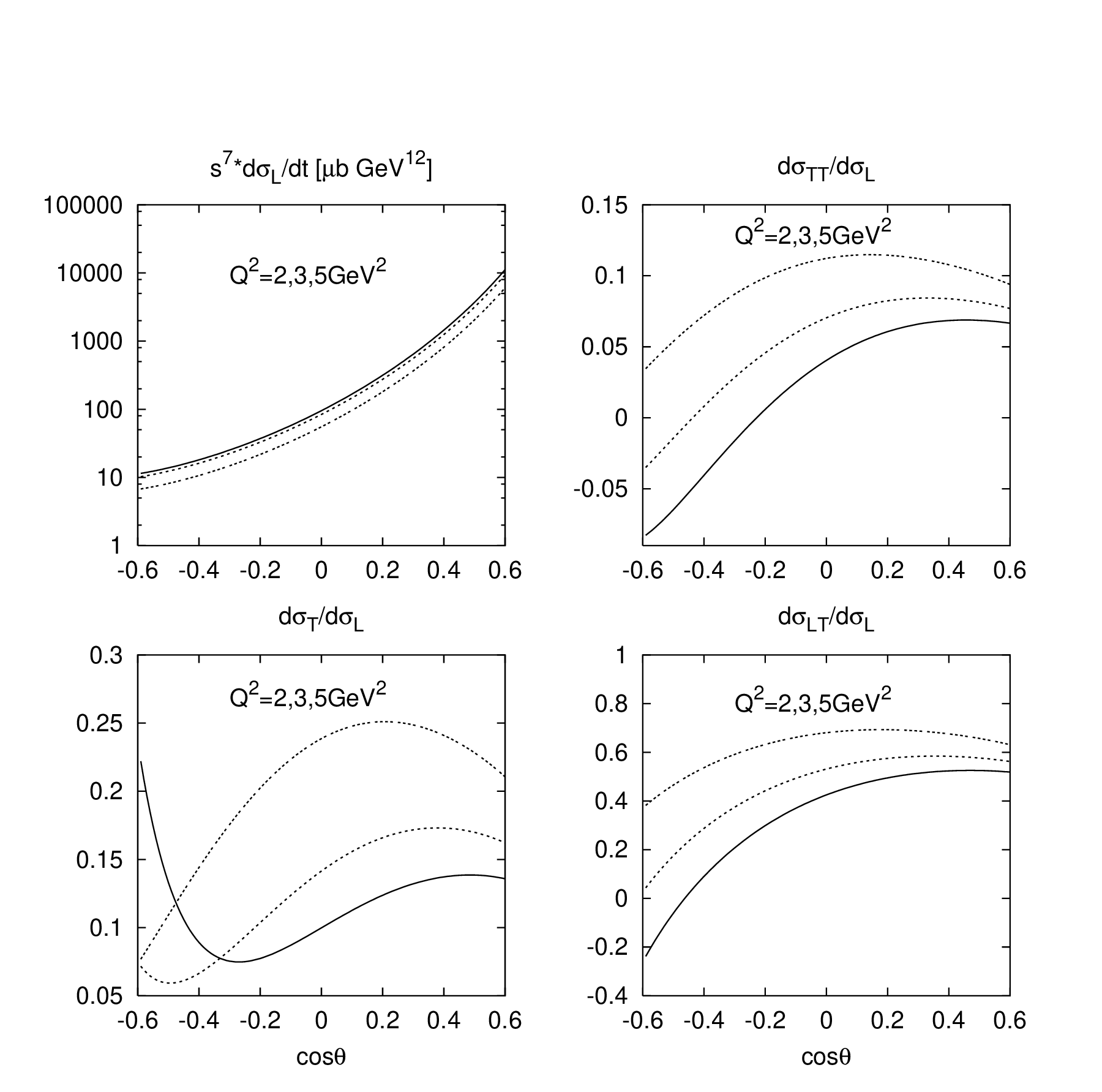,width=10.0cm,angle=-90}\\
\psfig{file=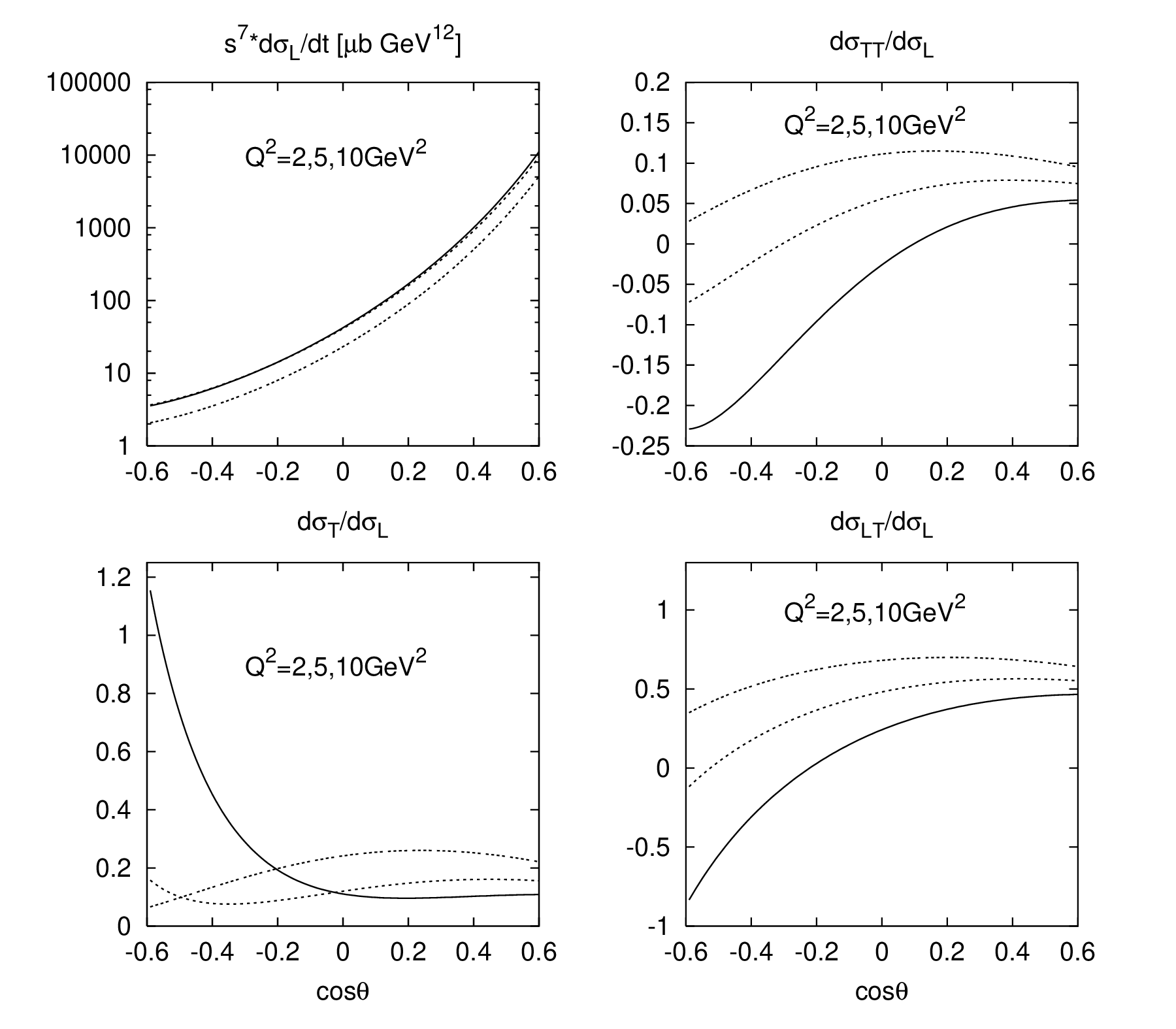,width=8.8cm,angle=-90}
\end{center}
\caption{The partial cross sections for electroproduction of $\rho^0$
         mesons versus $\cos{\theta}$ at $s=20\,\gev^2$ and
         $Q^2=2,3,5\,\gev^2$ (top) and at $s=40\,\gev^2$, 
         $Q^2=2,5,10\,\gev^2$ (bottom), plotted as solid, dotted and 
         dashed lines, respectively.}
\label{fig:9}
\end{figure}

Results for the partial cross sections for electroproduction of
$\rho^0$-mesons are shown in Fig.\ \ref{fig:9} (at $s=20\,\gev^2$ and 
$40\,\gev^2$). We see that the longitudinal cross section is dominant    
in the region of small $|\cos{\theta}|$. For larger values of 
$|\cos{\theta}|$ the transverse cross section as well as the 
longitudinal-transverse interference become sizeable. 

\begin{figure}[t]
\begin{center}
\psfig{file=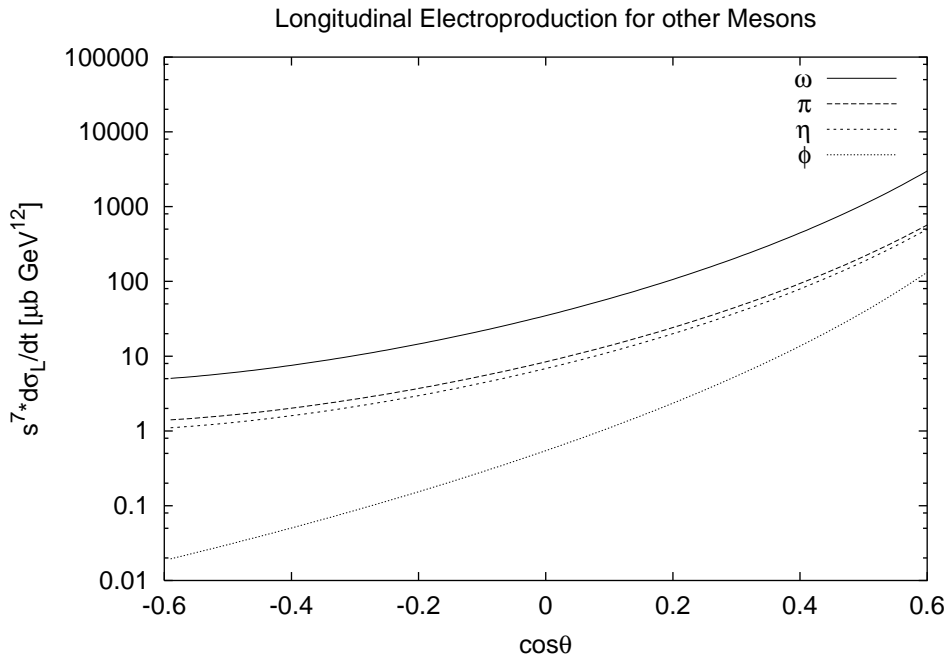,width=9cm,angle=-90}
\end{center}
\caption{The longitudinal cross section, scaled by $s^7$, for
         electroproduction of $\pi^0$, $\eta$, $\omega$ and $\phi$
         mesons versus $\cos{\theta}$
         at $s=20\,\gev^2$ and $Q^2=3\,\gev^2$.}
\label{fig:11}
\end{figure}

In order to demonstrate the relative magnitude of the production cross
sections for various mesons we display predictions for the scaled 
longitudinal cross sections at $s=20\,\gev^2$ and $Q^2=3\,\gev^2$ in 
Fig.\ \ref{fig:11}. The gluonic contributions to the cross sections 
for vector mesons are generally small in the kinematical region of 
interest. This is obvious from the relative strength of the quark and
gluon form factors, see Fig.\ \ref{fig:6}. The exceptional case is the
$\phi$ meson. The only quark form factor contributing, $R_V^{Vs}$, is
very small and the gluonic contribution therefore dominates the
production of $\phi$ mesons. Since the form factor $R^{Ps}_A$ is zero 
in the model (see Eq.\ (\ref{strange})) the ratio of
longitudinal cross sections for the production of $\eta'$ and $\eta$ 
mesons is given by the square of the tangent of the pseudoscalar meson 
mixing angle, see (\ref{mixing}).

Finally we comment on the accuracy of our calculation. As we said
repeatedly, large momentum transfer and photon virtualities larger
than about $2-3\gev^2$ are required for the dominance of the soft
mechanism. However, the momentum transfer  should not be too large
since then the pure perturbative contribution \cite{bro80,far} will
become important. The onset of the perturbative regime is expected to
be beyond $t\simeq -100 \gev^2$ \cite{DFJK1}. Even in the soft regime
one has to be aware of corrections. For instance, the lowest-order,
leading twist perturbative formation of the mesons may be subject to
substantial corrections of perturbative and/or soft origin. These may
give rise to a $\kappa$-factor in the normalization of the
electroproduction cross section as is known from the Drell-Yan process
and, according to Martin et al. \cite{mar96}, is required in DVEM at
least at small $Q^2/s$. 
\section{Summary}
In the present work electroproduction of flavor neutral pseudoscalar
and longitudinally polarized vector mesons at large $s$, $-t$ and $-u$
is investigated. The photon virtuality is not considered as a large
scale; therefore the limit $Q^2\to 0$ is included in the investigation. 
This study of electroproduction is complementary to the case  
of DVEM where $Q^2$ is large and $-t$ small. Based on the central 
assumption of the dominance of small parton virtualities and small 
intrinsic transverse momenta in the proton's light-cone wave function 
we have shown that, like in Compton scattering \cite{DFJK1}, the 
electroproduction amplitudes factorize in hard parton-level
subprocess amplitudes and soft proton matrix elements described by the 
same type of form factors as appear in Compton scattering. These form 
factors represent $1/x$-moments of SPDs. The soft mechanism bears
resemblance to the dynamics controlling DVEM  
\cite{rad96,CFS,man98,bro94,mar96} in many respects: The same 
parton-level subprocesses occur, the longitudinal cross section
dominates (if $|\cos{\theta}|$ is small and $Q^2$ not too small) and 
the soft information on the proton is encoded in SPDs. Different is
that, in the large momentum transfer region, a symmetric frame with
zero skewedness can be chosen which entails the formation of
$1/x$-moments of the SPDs, i.e.\ the appearance of new form factors. 
For asymptotically large momentum transfer the perturbative
contribution \cite{bro80} will take the lead, the soft contribution, 
discussed here, then presents a power correction to it. We emphasize 
that the dimensional counting rule behaviour, i.e.\ $s^7$-scaling,
approximately holds for photoproduction of mesons in the soft physics 
approach for a limited range of energy.

The new form factors, characteristic of large momentum transfer Compton
scattering and electroproduction of mesons, can in principle be
extracted from experiment by Rosenbluth-type separations
\cite{nathan}. Their measurements would provide information on the
large momentum transfer behavior of the proton SPDs and would allow to
test models for them. Moreover, the experimental verification of their
energy independence would constitute a severe test of the soft physics
approach.

Based on a light-cone wave function overlap model for the form
factors we have presented detailed predictions for
electroproduction of pseudoscalar and longitudinally polarized vector
mesons at moderately large photon virtuality. Although the soft
physics approach also applies to large momentum transfer
photoproduction of mesons it seems - as judged on the basis of the 
present data - that the contributions from the hadronic component of 
the photon dominate these reactions up to rather high energies. 
The kinematical region in which the soft physics approach is
applicable to electroproduction, is accessible to experiments at the
upgraded TJlab and at the proposed ELFE accelerator and EPIC collider. 
The measurement of large momentum transfer electroproduction of mesons
is certainly difficult but seems feasible.
We have not discussed electroproduction of flavored mesons and of
transversally polarized vector mesons in this work. These processes
involve flavor or helicity non-diagonal SPDs which are not directly 
related to those appearing in the processes we have investigated. As
in DVEM \cite{die99,man00}, higher twist dynamics plays an important
role in electroproduction of transversally polarized vector mesons
since the leading twist, lowest order subprocess amplitudes are zero
in this case.
 
\section{Acknowledgment}
We would like to thank Markus Diehl, Thorsten Feldmann and Rainer
Jakob for many stimulating discussions.
H.W.\ Huang thanks the Deutsche Forschungsgemeinschaft for support.

\end{document}